\documentclass[prd,preprintnumbers,nofootinbib]{revtex4} 

\usepackage{amsmath}
\usepackage{epsfig,epsf}
\usepackage{graphicx}
\usepackage{verbatim}
\usepackage{epstopdf}

\def\k{{\mathbf k}}

\def\p{{\mathbf p}}

\def\k{{\mathbf k}}
\def\x{{\mathbf x}}

\def\r{{\mathbf r}}

\def\b{{\mathbf b}}

\def\bq{\bar{q}}

\newcommand{\beq}{\begin{eqnarray}}
\newcommand{\eeq}{\end{eqnarray}}
\newcommand{\be}{\begin{eqnarray*}}
\newcommand{\ee}{\end{eqnarray*}}
\DeclareMathOperator{\tr}{tr}

\def\be{\begin{equation}}
\def\ee{\end{equation}}
\def\bea{\begin{eqnarray}}
\def\eea{\end{eqnarray}}

\def\eq#1{{Eq.~(\ref{#1})}}
\def\fig#1{{Fig.~\ref{#1}}}

\begin{document}

\title{Diffractive Dijet Production in Deep Inelastic Scattering and Photon-Hadron Collisions in the Color Glass Condensate}

\author{Tolga Altinoluk$^1$, N\'estor Armesto$^1$, Guillaume Beuf$^{2,3}$ and  Amir H. Rezaeian$^{4,5}$}

\affiliation{$^1$ Departamento de F\'{i}�sica de Part\'{i}culas and IGFAE, Universidade de Santiago de Compostela, 15706 Santiago de Compostela, Galicia-Spain\\
$^2$ Department of Physics, Ben-Gurion University of the Negev,
Beer Sheva 84105, Israel\\
$^3$ European Centre for Theoretical Studies in Nuclear Physics and Related Areas (ECT*)
and Fondazione Bruno Kessler, Strada delle Tabarelle 286,
 I-38123 Villazzano (TN), Italy\\
$^4$ Departamento de F\'{\i}sica, Universidad T\'ecnica Federico Santa Mar\'{\i}a,
Avda. Espa$\tilde{\rm n}$a 1680, Casilla 110-V, Valpara\'{\i}so, Chile\\
$^5$ Centro Cient\'{\i}fico Tecnol\'ogico de Valpara\'{\i}so (CCTVal),
Universidad T\'ecnica Federico Santa Mar\'{\i}a,
Casilla 110-V, Valpara\'{\i}so, Chile
}

\date{\today}

\begin{abstract}
We study exclusive  dijet production in coherent diffractive processes in deep inelastic scattering and real (and virtual) photon-hadron ($\gamma^{(*)}$-h) collisions in the Color Glass Condensate formalism at leading order. We show that the diffractive dijet cross section is sensitive to the color-dipole orientation in the transverse plane, and is a good probe of possible correlations between the $q\bar{q}$-dipole transverse separation vector $\r$ and the dipole impact parameter $\b$. We also investigate the diffractive dijet azimuthal angle correlations and $t$-distributions in $\gamma^{(*)}$-h collisions and show that they are sensitive to gluon saturation effects in the small-$x$ region. In particular, we show that the $t$-distribution of diffractive dijet photo-production off a proton target exhibits a dip-type structure in the saturation region. This effect is similar to diffractive vector meson production. Besides, at variance with the inclusive case, the effect of saturation leads to stronger azimuthal correlations between the jets.

\end{abstract}

\maketitle

\section{Introduction}
\label{intro}
Diffractive production provides a rich testing ground of many novel properties of Quantum Chromodynamics (QCD), see for example \cite{all-diff}. 
In particular, the diffractive deep inelastic scattering (DIS) offers an opportunity to explore the interesting transition from hard to soft physics, via purely perturbative calculations \cite{Wusthoff:1999cr,Wolf:2009jm}. 
The analysis of diffractive processes in  real and virtual photon-hadron ($\gamma^{(*)}$-h) collisions is one of the largest sources of information to explore the behavior of the strong interaction at high energies, for a review see Refs.\,\cite{Wusthoff:1999cr,Wolf:2009jm}.  For example, by measuring the squared momentum transfer $t$, one can study the transverse spatial distribution of the gluons (e.g. an eventual  hot-spot structure) in the hadron wave function that cannot be probed in inclusive processes.

Inclusive diffractive DIS has been extensively studied by the H1 and ZEUS collaborations at HERA, for the most recent analysis see Refs.\,\cite{Aaron:2012ad,Chekanov:2008fh,Aaron:2012hua,h1-dijet}. On the theoretical side, it was shown that a factorization theorem exists for inclusive diffractive electron-proton DIS within the collinear framework \cite{Collins:1997sr} and, thereby, one can extract the diffractive parton densities \cite{Capella:1995pn,Bartels:1998ea,Martin:2005hd}.  This is in contrast to hard processes in diffractive hadron-hadron scattering (like diffractive Drell-Yan) where such factorization fails \cite{collins-dy}. 
Diffractive jet production was later proposed to test further the collinear factorization and, thus, to provide complementary information about the underlying dynamics of high-energy $\gamma^{(*)}$-h collisions  \cite{Bartels:1996ne,Bartels:1996tc,Bartels:1999tn,Bartels:2002ri,Braun:2005rg}. However, the detailed analyses performed in the collinear framework indicate a {sizable excess of the next-to-leading order (NLO) predictions over the experimental data \cite{Adloff:2000qi,Aktas:2006up,Andreev:2014yra,Andreev:2015cwa,Chekanov:2007rh,Aaron:2010su,Aaron:2011mp,Abramowicz:2015vnu} that is attributed to the existence of absorptive corrections and collinear factorization breaking \cite{Klasen:2008ah,Kaidalov:2003xf}, see also Ref.\,\cite{h1-dijet}. Therefore, it is important to consider schemes alternative to the collinear factorization where diffractive jet production can be computed.

The collinear factorization is expected to be valid when partonic system in the hadron is in the dilute regime i.e. when radiation from each parton can be considered as independent which is expected to be the physical situation at large enough scales (photon virtuality and jet transverse energy). On the other hand, the Color Glass Condensate (CGC) offers a framework for computing such observable when the density of partons is high, for recent reviews see Refs.\,\cite{Gelis:2010nm,Albacete:2014fwa}. In the CGC formalism, the standard quantum evolution equations with large logarithms of $1/x$ resummed, lead to a scenario in which the occupancy of the slow modes in the target hadron is so high that they can be treated classically, with the fast modes considered as sources. The corresponding renormalisation
group equations which govern the separation of the fast and slow models, known in the limit of scattering of a dilute probe on a dense hadron, are the so-called Balitsky--Jalilian-Marian-Iancu-McLerran-Weigert-Leonidov-Kovner (B-JIMWLK) hierarchy of equations \cite{jimwlk} or, in the large $N_c$ limit, the Balitsky-Kovchegov (BK) equation \cite{bk}, recently computed up to NLO accuracy \cite{Balitsky:2008zza,Kovner:2013ona}.

In this paper, we consider exclusive  dijet production in coherent diffractive processes, i.e. production of just two fully reconstructed jets with the target hadron remaining intact, in dilute-dense scatterings such as real and virtual photon-hadron (or photon-nucleus) collisions in the CGC approach at the lowest order.  Previous calculations performed under this framework have mainly focused on the inclusive dijet production \cite{in-dijet1,Mueller:2012uf,Mueller:2013wwa}, which are calculations involving a different color structure and color averaging than the process we are interested in. The cross section of inclusive dijet production in DIS depends on two-point and four-point functions, and therefore involving Weizs\"acker-Williams gluon distribution in the correlation limit. In drastic contrast, here we show that diffractive dijet process in DIS only involves the dipole gluon distribution (and two-point functions), see also Refs.\,\cite{dijet2,dijet3}. We show that the diffractive dijet cross section in DIS can be written in terms of a convolution of two forward dipole amplitudes in which the color-dipole orientation in the transverse plane  becomes a crucial ingredient. In this sense, diffractive dijet production in DIS is a sensitive probe of the color-dipole amplitude orientation, and provides useful information about unknown correlations between the impact parameter $\b$ and the $q\bar{q}$-dipole transverse separation vector $\r$. This is a very important feature of our calculation that, to our knowledge, does not appear in previous CGC-based calculations of other processes.

We also investigate the diffractive dijet azimuthal angle correlations and $t$-distributions in $\gamma^{(*)}$-h collisions and show that they are sensitive to gluon saturation effects in the small-$x$ kinematic region. In particular, we show that the $t$-distribution of diffractive dijet photo-production off a proton target exhibits a dip-type structure in the saturation region. A similar behavior was also recently reported to exist for diffractive photo-production of vector mesons \cite{na}. Note that, in  contrast to diffractive vector meson production, diffractive dijet production is free from final-state hadronization effects. Therefore, our calculation here provides a strong indication that the emergence of a dip-type structure is a universal feature of diffractive production in the small-$x$ region, and it does not depend on the details of the final-state particle wave functions. We provide various predictions which can further test this idea at the Large Hadron Collider (LHC) and future DIS experiments.

This paper is organised as follows: in Sec. II the theoretical calculation is performed, with the several technical steps detailed. In Sec. III, numerical results are presented for some kinematic situations that can be achieved at the LHC and in future electron-proton/nucleus colliders \cite{Accardi:2012qut,AbelleiraFernandez:2012cc,fcc}. Finally, in Sec. \ref{conclu} our results are summarized and discussed.   In the Appendix we consider a simple dipole model with polarization in order to explicitly show that the diffractive dijet cross section is sensitive to the color-dipole orientation. 

\section{Theoretical formalism:  setup and formulas}

We consider the process  $\gamma^*+ p(A)\rightarrow q+\bar{q}+p(A)$ in the CGC approach, as illustrated in Fig. \ref{fig1}.
\begin{figure} [hbt]
\includegraphics[scale=0.5]{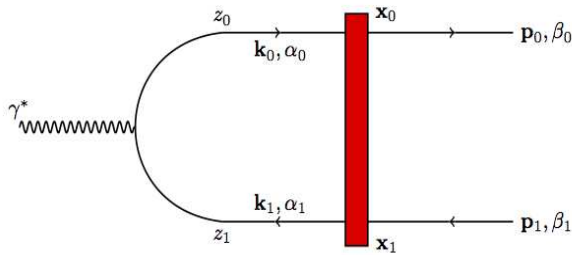}
\caption{Diagram for lowest-order of diffractive dijet production in DIS in the CGC framework.}
\label{fig1}
\end{figure}
%

At leading order, the incoming wave-function of the virtual photon can be written in the mixed space representation as 
\beq
|\gamma^*_{\rm phys}\rangle
=\sum_{\rm q\bar{q}\; states} \delta_{\alpha_0 \alpha_1} \; \Psi(z_0,z_1,\x_0,\x_1)\; | q_{\alpha_0} \bar{q}_{\alpha_1}\rangle_{\rm mixed}.
\label{gammawavef}
\eeq
 Here $z_i$ stands for the longitudinal momentum fraction of the quark and antiquark with respect to the longitudinal momentum of the virtual photon and is defined as $z_i=k^+_i/q^+=p^+_i/q^+$. By $q^{\mu}$ we denote the four-momenta of the incoming virtual photon with virtuality $q^2=-Q^2$. Two-dimensional vectors in transverse space are written in boldface.
Finally, $\alpha_i$ are the color indices of the quark and antiquark before the interaction with the target whereas $\beta_i$ will be the color indices after the interaction\footnote{We follow the same conventions as in \cite{Beuf:2011xd}.}. The $q\bar{q}$ state is obtained via the action of the quark and antiquark creation operators on the vacuum state
\beq
| q_{\alpha_0} \bar{q}_{\alpha_1}\rangle_{\rm mixed}=b^{\dagger}(\x_0,z_0,\alpha_0,h_0)d^{\dagger}(\x_1,z_1,\alpha_1,h_1)|0\rangle, 
\eeq 
with $h_i$ being the helicity.

For a dense enough target or at low Bjorken-$x$, the occupation number of soft gluons becomes 
large enough, so that the state of the target can be better described in terms of semi-classical gluon fields  than of Fock
states with a finite number of individual gluons.  In the limit of infinite boost, the target
field strength is subject to Lorentz contraction and can be considered as a shock-wave field. The incoming wave function of the virtual photon interacts with the target, for which we will employ the eikonal approximation. In this approximation, the transverse position $\x_{0,1}$ of partons
stays constant during their infinitely fast interaction with the infinitely thin shock-wave, and each parton in the incoming wave function picks only an eikonal phase during the interaction with the target \cite{Bjorken:1970ah}. The outgoing wave function after the interaction, in mixed space representation, reads
\beq 
\hat{S}|\gamma^*_{\rm phys}\rangle&=&\sum_{\rm q\bar{q}\; states} \left[ U_F(\x_0)U^{\dagger}_F(\x_1)\right]_{\beta_0 \beta_1}\; \Psi_{q\bq}(z_0,z_1,\x_0,\x_1) \; |q_{\beta_0} \bar{q}_{\beta_1}\rangle_{\rm mixed} \, ,
\eeq
where $U_F(\x)$ is the Wilson line for the propagation of a high energy parton in the fundamental representation of $SU(N_c)$ at transverse position $\x$. The $q\bar{q}$ states that are present both in the incoming and outgoing wave functions are written in the mixed space representation. However, one can write it in momentum space representation as well via a Fourier transformation, reading
%
\beq 
\hat{S}|\gamma^*_{\rm phys}\rangle&=&\sum_{\rm q\bar{q}\; states}\int \frac{d^2 \x_0}{2\pi} \int \frac{d^2 \x_1}{2\pi} e^{-i\p_0\cdot \x_0} e^{-i\p_1\cdot \x_1}   \left[ U_F(\x_0)U^{\dagger}_F(\x_1)\right]_{\beta_0 \beta_1}\; \Psi_{q\bq}(z_0,z_1,\x_0,\x_1) \; |q_{\beta_0} \bar{q}_{\beta_1}\rangle_{\rm mom} \, .
\eeq
Within the CGC effective theory, the $S$-matrix of the process is given by the overlap of the incoming and outgoing wave functions:
\beq
S_{q\bq\leftarrow\gamma}&=&\langle \bq_{\beta_1}(\p_1,z_1,h_1)\; q_{\beta_0}(\p_0,z_0,h_0) | \hat{S} | \gamma^*_{\rm phys} \rangle\, .
\eeq
%
Using the momentum space expressions of the incoming and outgoing wave functions, one gets the explicit form of the $S$-matrix of the process:
\beq
S_{q\bq\leftarrow\gamma}=\int \frac{d^2 \x_0}{2\pi} \int \frac{d^2 \x_1}{2\pi} e^{-i\p_0\cdot \x_0} e^{-i\p_1\cdot \x_1}   \left[ U_F(\x_0)U^{\dagger}_F(\x_1)\right]_{\beta_0 \beta_1}\; \Psi_{q\bq}(z_0,z_1,\x_0,\x_1) \, .
\eeq
The corresponding forward scattering amplitude, ${\cal M}_{q\bq\leftarrow\gamma}$, reads
\beq
2\pi (2q^+) \delta(k^+_0+k^+_1-q^+){\cal M}_{q\bq\leftarrow\gamma}&=&\int \frac{d^2 \x_0}{2\pi} \int \frac{d^2 \x_1}{2\pi} e^{-i\p_0\cdot \x_0} e^{-i\p_1\cdot \x_1}  \left( \left[ U_F(\x_0)U^{\dagger}_F(\x_1)\right]_{\beta_0 \beta_1} -\delta_{\beta_0 \beta_1}\right) \nonumber\\
&&\hspace{2.5cm}\times \Psi_{q\bq}(z_0,z_1,\x_0,\x_1) \, .
\eeq
The main difference between diffractive and inclusive productions is that for diffractive production one has to perform averaging over the color sources of the target at the amplitude level $\sigma\propto |\langle \mathcal{M} \rangle_{C}|^2$, while for inclusive production averaging over the color is performed at the cross section level  $\sigma\propto \langle |\mathcal{M}|^2 \rangle_{C}$, where $\langle \cdots \rangle_C$ stands for the average on the target color configurations. The diffractive dijet cross section can then be written in terms of the forward scattering amplitude as
\beq
(2\pi)^6\;2p^+_0\;2p^+_1\frac{d\sigma^{\rm diff \; dijet}}{dp^+_0dp^+_1d^2\p_0d^2\p_1}=\sum_{\beta_i h_i}(2q^+)\;2\pi\;\delta(k^+_0+k^+_1-q^+) \bigl|\langle{\cal M}_{q\bq\leftarrow\gamma}\rangle_C\bigr|^2,
\eeq
with 
\beq
(2q^+)\;2\pi\;\delta(k^+_0+k^+_1-q^+) \langle{\cal M}_{q\bq\leftarrow\gamma}\rangle_C = \int \frac{d^2 \x_0}{2\pi} \int \frac{d^2 \x_1}{2\pi} e^{-i\p_0\cdot \x_0} e^{-i\p_1\cdot \x_1} \delta_{\beta_0 \beta_1}\left( S_{01} - \mathbf{1}\right) \Psi_{q\bq}(z_0,z_1,\x_0,\x_1), 
\eeq
 where $S_{01}$ is the usual dipole operator in the fundamental representation:
\beq
S_{01}=\Big\langle \frac{1}{N_c}\tr \left[ U_F(\x_0)U^{\dagger}_F(\x_1)\right] \Big\rangle_C\,.
\eeq
The function $\Psi$ appearing in the definition of the $q\bar{q}$ states can be written explicitly \cite{Beuf:2011xd} as 
\beq
\Psi_{q\bq}(\x_0,\x_1,z_o,z_1,h_0,h_1)=(2\pi)^2\;2ee_f\;\sqrt{z_0z_1}\;\delta(z_0+z_1-1)\;\Phi^{\rm LO}_{\rm T,L}(\x_0,\x_1,z_0,z_1,(h_0),\lambda)\;\delta_{h_0,-h_1}, 
\label{psi_vs_phi}
\eeq
where $ee_f$ is the electric charge of quark $f$ and the transverse (T) and longitudinal (L) contributions read
\beq
\Phi^{\rm LO}_{\rm T}(\x_0,\x_1,z_0,z_1,h_0,\lambda)&=&i[z_1-z_0-(2h_0)\lambda] \; \frac{\epsilon_{\lambda}\cdot\x_{01}}{x^2_{01}} \; Q\sqrt{z_0z_1x^2_{01}} \; K_1\left(Q\sqrt{z_0z_1x^2_{01}}\right),
\label{phiT}
\\
\Phi^{\rm LO}_{\rm L}(\x_0,\x_1,z_0,z_1)&=& -2z_0z_1 \; Q \; K_0\left(Q\sqrt{z_0z_1x^2_{01}}\right),
\label{phiL}
\eeq
where we used the notations $\x_{ij}=\x_i-\x_j$, $x_{ij}=|\x_{ij}|$. From now on,  $\lambda$ is the polarization of the (transverse) photon, and $K_0$ and $K_1$ are modified Bessel functions of the second kind.

By using Eq. \eqref{psi_vs_phi}, the forward scattering amplitude can be written as 
\beq
{\cal M}_{q\bq\leftarrow\gamma}&=&(2\pi)  ee_f \sqrt{z_0z_1} \int \frac{d^2\x_0}{2\pi}\int \frac{d^2\x_1}{2\pi} e^{-i\p_0\cdot\x_0}e^{-i\p_1\cdot\x_1}\left( \left[ U_F(\x_0) U^{\dagger}_F(\x_1)\right]_{\beta_0\beta_1}\!\!-\delta_{\beta_0\beta_1}  \right)\nonumber\\
&&\hspace{4.7cm}\times \Phi^{\rm LO}_{\rm T,L}(\x_0,\x_1,z_0,z_1,(h_0),\lambda)\;\delta_{h_0,-h_1} \, .
\eeq
Finally, after averaging over the target color fields,  the forward scattering amplitude reads
\beq
\langle {\cal M}_{q\bq\leftarrow\gamma}\rangle_C=(2\pi) \; ee_f \; \sqrt{z_0z_1} \delta_{\beta_0\beta_1} \delta_{h_0,-h_1} \int \frac{d^2\x_0}{2\pi}\int \frac{d^2\x_1}{2\pi} e^{-i\p_0\cdot\x_0}e^{-i\p_1\cdot\x_1}(-{\cal N}_{01})\Phi^{\rm LO}_{\rm T,L}(\x_0,\x_1,z_0,z_1,(h_0),\lambda),
\eeq
where ${\cal N}_{01}$ is the dipole amplitude which is given as $S_{01}=1-{\cal N}_{01}$.

Thus, the diffractive cross section is given by 
\beq
&&(2\pi)^6\; 2p^+_0\;2p^+_1\left( \frac{d\sigma^{\rm diff \;dijet}_{\rm T,L}}{dp^+_0dp^+_1d^2\p_0d^2\p_1}\right)=(2q^+)2\pi\delta(p^+_0+p^+_1-q^+)\sum_{\beta_i,h_i,f}\big|\langle {\cal M}_{q\bq\leftarrow\gamma}\rangle_C\big|^2\nonumber\\
&&=4\alpha_{em}N_c(2\pi)^4\delta(z_0+z_1-1)\left( \sum_{f}e^2_f\right)z_0z_1\int \frac{d^2\x_0}{2\pi}\int \frac{d^2\x_1}{2\pi} e^{-i\p_0\cdot\x_0}e^{-i\p_1\cdot\x_1}{\cal N}_{01}\nonumber\\
&&\times\int \frac{d^2\x'_0}{2\pi}\int \frac{d^2\x'_1}{2\pi} e^{-i\p_0\cdot\x'_0}e^{-i\p_1\cdot\x'_1}{\cal N}_{1'0'}\;\Phi^{\rm LO}_{\rm T,L}(\x_0,\x_1,z_0,z_1,(h_0),\lambda)\;\Phi^{\rm LO}_{\rm T,L}(\x'_0,\x'_1,z_0,z_1,(h_0),\lambda)^*,
\eeq
{\bf where $\alpha_{em}=e^2/(4\pi)$ is the} fine structure constant.
For the case of longitudinal photon, using Eq.\,\eqref{phiL} we obtain
\beq
\sum_{h_0,h_1}\delta_{h_0,-h_1}\Phi^{\rm LO}_{\rm L}(\x_0,\x_1,z_0,z_1)\Phi^{\rm LO}_{\rm L}(\x'_0,\x'_1,z_0,z_1)^*=8\;z^2_0z^2_1\;Q^2\;K_0\left(Q\sqrt{z_0z_1x^2_{01}}\right)K_0\left(Q\sqrt{z_0z_1x^2_{1'0'}}\right).
\eeq 
In the same fashion, for the case of transverse photon, using Eq.\,\eqref{phiT} we obtain
\beq
&&\sum_{h_0,h_1}\delta_{h_0,-h_1}\Phi^{\rm LO}_{\rm T}(\x_0,\x_1,z_0,z_1,h_0,\lambda)\Phi^{\rm LO}_{\rm T}(\x'_0,\x'_1,z_0,z_1,h_0,\lambda)^*= 4[z^2_0+z^2_1]\left(\frac{\epsilon_{\lambda}\cdot\x_{01}}{x^2_{01}}\right)\left(\frac{\epsilon^*_{\lambda}\cdot\x_{1'0'}}{x^2_{1'0'}}\right)\nonumber\\
&&\times\left[Q\sqrt{z_0z_1x^2_{01}}K_1\left(Q\sqrt{z_0z_1x^2_{01}}\right)\right]\; \left[Q\sqrt{z_0z_1x^2_{1'0'}}K_1\left(Q\sqrt{z_0z_1x^2_{1'0'}}\right)\right],
\eeq
and averaging over $\lambda$ results in 
\beq
&&\frac{1}{2}\sum_{\lambda}\left(\sum_{h_0,h_1}\delta_{h_0,-h_1}\Phi^{\rm LO}_{\rm T}(\x_0,\x_1,z_0,z_1,h_0,\lambda)\Phi^{\rm LO}_{\rm T}(\x'_0,\x'_1,z_0,z_1,h_0,\lambda)^*\right)=2 [z^2_0+z^2_1]\frac{\x_{01}\cdot\x_{1'0'}}{x^2_{01}x^2_{1'0'}}\\\nonumber
&&\times\;\left[Q\sqrt{z_0z_1x^2_{01}}K_1\left(Q\sqrt{z_0z_1x^2_{01}}\right)\right]\; \left[Q\sqrt{z_0z_1x^2_{1'0'}}K_1\left(Q\sqrt{z_0z_1x^2_{1'0'}}\right)\right].
\eeq

Let us define $\varepsilon^2=z_0z_1Q^2$, $\r=\x_{01}$, $\r'=\x_{1'0'}$, $\b=(\x_0+\x_1)/2$ and $\b'=(\x'_0+\x'_1)/2$, with $\r$ (and $\r'$) and $\b$ (and $\b'$) being $q\bar{q}$-dipole transverse separation vector and dipole impact parameter in the amplitude (and in its complex-conjugate), respectively.  Then, the final expressions for the diffractive dijet cross section can be written in the following factorization form for the longitudinal component:
\beq
&&p^+_0p^+_1\left( \frac{d\sigma^{\rm diff \; dijet}_{\rm L}}{dp^+_0dp^+_1d^2\p_0d^2\p_1}\right)= (2\pi)^2\delta(z_0+z_1-1)N_c\alpha_{em}\sum_fe^2_f\;z_0z_1\int \frac{d^2\r}{(2\pi)^2}\int \frac{d^2\r'}{(2\pi)^2}\int \frac{d^2\b}{(2\pi)^2}\int \frac{d^2\b'}{(2\pi)^2}\nonumber\\
&&\times \; e^{-i(\b-\b')\cdot(\p_0+\p_1)}e^{-i(\r-\r')\cdot(\p_0-\p_1)/2}{\cal N}(\r,\b)\;{\cal N}(\r',\b')\;8z_0z_1\varepsilon^2\;K_0(\varepsilon|\r|)\;K_0(\varepsilon|\r'|),
\label{Lf1}
\eeq
 and for the  transverse component 
\beq
&&p^+_0p^+_1\left( \frac{d\sigma^{\rm diff \; dijet}_{\rm T}}{dp^+_0dp^+_1d^2\p_0d^2\p_1}\right)= (2\pi)^2\delta(z_0+z_1-1)N_c\alpha_{em}\sum_fe^2_f\;z_0z_1\int \frac{d^2\r}{(2\pi)^2}\int \frac{d^2\r'}{(2\pi)^2}\int \frac{d^2\b}{(2\pi)^2}\int \frac{d^2\b'}{(2\pi)^2}\nonumber\\
&&\times \; e^{-i(\b-\b')\cdot(\p_0+\p_1)}e^{-i(\r-\r')\cdot(\p_0-\p_1)/2}{\cal N}(\r,\b)\;{\cal N}(\r',\b')\; 
2[z^2_0+z^2_1]\;\frac{\r\cdot\r'}{r^2{r'}^2}\; \left[\varepsilon |\r|K_1(\varepsilon |\r|)\right]\;\left[\varepsilon |\r'|K_1(\varepsilon |\r'|)\right].
\label{Tf1}
\eeq
 It is obvious from the  expressions \eqref{Lf1} and \eqref{Tf1} that the cross section of diffractive dijet production explicitly depends on the color dipole orientation, namely on the azimuthal angle between the $q\bar{q}$-dipole transverse separation vector $\r$ and the dipole impact parameter $\b$. This is one of the most salient features of our calculations. Nevertheless, we will not exploit it in the rest of the paper due to the fact that none of the existing models for the dipole amplitude consider it. In the next subsection, we show that one can drastically simplify these expressions by assuming azimuthal symmetry in the color-dipole amplitude. 

\subsection{Azimuthal Symmetry}

The expressions for the diffractive cross sections Eqs.\,(\ref{Lf1},\ref{Tf1}), can be further simplified using azimuthal symmetry and the independence on the relative angle between vectors $\r$ and $\b$. 
We denote by  $\theta_r, \theta_b,  \theta_+$ and $\theta_-$ the angles of vectors $\r$, $\b$, $(\p_0+\p_1)$ and $(\p_0-\p_1)$ with respect to a reference vector, respectively.  Let us first assume that ${\cal N}(\r,\b)={\cal N}(r,b,\theta_r-\theta_b)$. By using this symmetry for the case of the longitudinal photon, one can simplify 
Eq.\,\eqref{Lf1} by noting that}
\beq \label{azy}
&&\int \frac{d^2\r}{(2\pi)^2}\int \frac{d^2\b}{(2\pi)^2} e^{-i\b\cdot(\p_0+\p_1)}e^{-i\r\cdot(\p_0-\p_1)/2}{\cal N}(\r,\b)K_0(\varepsilon|\r|)=\int_0^{+\infty}\frac{dr}{2\pi}r\int_0^{+\infty}\frac{db}{2\pi}b\int_0^{2\pi}\frac{d\theta_r}{2\pi}\int_0^{2\pi}\frac{d\theta_b}{2\pi}\nonumber\\
&&\times \; e^{-ib|\p_0+\p_1|\cos(\theta_b-\theta_+)}e^{-i\frac{r}{2}|\p_0-\p_1|\cos(\theta_r-\theta_-)}{\cal N}(r,b,\theta_r-\theta_b)K_0(\varepsilon|\r|).
\eeq
Now if we assume ${\cal N}(r,b,\theta_r-\theta_b)={\cal N}(r,b)$, namely  that the dipole amplitude does not depend on the angle between the dipole transverse separation vector and the impact parameter, one can further simplify this expression by analytically performing the integral over the angles. Therefore, we obtain, 
\beq
\int \frac{d^2\r}{(2\pi)^2}\int \frac{d^2\b}{(2\pi)^2} e^{-i\b\cdot(\p_0+\p_1)}e^{-i\r\cdot(\p_0-\p_1)/2}{\cal N}(\r,\b)K_0(\varepsilon|\r|)&=&
\int_0^{+\infty}\frac{dr}{2\pi}r\int_0^{+\infty}\frac{db}{2\pi}b \;J_0\left(b|\p_0+\p_1|\right)\nonumber\\
&\times& \; J_0\left(r\frac{|\p_0-\p_1|}{2}\right){\cal N}(r,b)K_0(\varepsilon r).
\eeq
Finally, the diffractive cross section for the longitudinal photon reads
\beq
p^+_0p^+_1\left( \frac{d\sigma^{\rm diff \; dijet}_{\rm L}}{dp^+_0dp^+_1d^2\p_0d^2\p_1}\right)&=&\delta(z_0+z_1-1)\;\frac{N_c\alpha_{em}}{(2\pi)^2}\; 8 \; z^2_0 \; z^2_1\; {\varepsilon}^2\nonumber\\
&\times&\left[\int_0^{+\infty}\!\!\!dr \;r \int_0^{+\infty}\!\!\!db \; b J_0\left(b|\p_0+\p_1|\right) \; J_0\left(r\frac{|\p_0-\p_1|}{2}\right){\cal N}(r,b)K_0(\varepsilon r)\right]^2.
\eeq

A similar simplification is valid also for the case of the transverse photon. Assuming the same azimuthal symmetry, it is again straightforward to realise that
\beq 
&&\int \frac{d^2\r}{(2\pi)^2}\int \frac{d^2\b}{(2\pi)^2} e^{-i\b\cdot(\p_0+\p_1)}e^{-i\r\cdot(\p_0-\p_1)/2}{\cal N}(\r,\b)\left(\varepsilon |\r| K_1(\varepsilon|\r|)\right)\frac{\r^j}{r^2}
= 2i\partial_{(\p^j_0-\p^j_1)}\int_0^{+\infty}\frac{dr}{2\pi}\frac{1}{r}\left[\varepsilon r K_1(\varepsilon r)\right]\nonumber\\
&&\times\int_0^{+\infty}\frac{db}{2\pi}b \int_0^{2\pi}\frac{d\theta_r}{2\pi}\int_0^{2\pi}\frac{d\theta_b }{2\pi}
e^{-ib|\p_0+\p_1|\cos(\theta_b-\theta_+)}e^{-i\frac{r}{2}|\p_0-\p_1|\cos(\theta_r-\theta_-)}{\cal N}(r,b,\theta_r-\theta_b) \, \label{azy2}
\eeq
Moreover,  assuming again ${\cal N}(r,b,\theta_r-\theta_b)={\cal N}(r,b)$, one gets 
\beq
&&\int \frac{d^2\r}{(2\pi)^2}\int \frac{d^2\b}{(2\pi)^2} e^{-i\b\cdot(\p_0+\p_1)}e^{-i\r\cdot(\p_0-\p_1)/2}{\cal N}(\r,\b)\left(\varepsilon |\r| K_1(\varepsilon|\r|)\right)\frac{\r^j}{r^2}
= 2i\partial_{(\p^j_0-\p^j_1)}\int_0^{+\infty}\frac{dr}{2\pi}\frac{1}{r}\left[\varepsilon r K_1(\varepsilon r)\right]\nonumber\\
&&\hspace{3cm}\times\int_0^{+\infty}\frac{db}{2\pi}b\; {\cal N}(r,b)\; J_0\left(b|\p_0+\p_1|\right)\; J_0\left(\frac{r}{2}|\p_0-\p_1|\right).
\eeq
Nothing that
\beq
\partial_{(\p^j_0-\p^j_1)}J_0\left(\frac{r}{2}|\p_0-\p_1|\right)=-\frac{r}{2}J_1\left(\frac{r}{2}|\p_0-\p_1|\right)\frac{(\p^j_0-\p^j_1)}{|\p_0-\p_1|}\ ,
\eeq
the diffractive cross section for the transverse photon finally reads
\beq \label{mless}
&&p^+_0p^+_1\left( \frac{d\sigma^{\rm diff \; dijet}_{\rm T}}{dp^+_0dp^+_1d^2\p_0d^2\p_1}\right)=\delta(z_0+z_1-1)\frac{N_c\alpha_{em}}{(2\pi)^2}\sum_f e^2_f\;2\;z_0z_1[z^2_0+z^2_1]\nonumber\\
&&\times\left[\int_0^{+\infty}dr\left(\varepsilon r K_1(\varepsilon r)\right)J_1\left(\frac{r}{2}|\p_0-\p_1|\right)\int_0^{+\infty}db \; b \; J_0(b|\p_0+\p_1|)\;{\cal N}(r,b) \right]^2.
\eeq

\subsection{Including quark mass}

Including the quark mass $m_f$ (for a given flavor $f$) in the cross section for the longitudinal photon is straightforward. It simply appears in $\varepsilon $ through the following substitution:
\beq
\varepsilon^2\rightarrow \varepsilon^2_f=z_0z_1Q^2+m^2_f\, .
\eeq
For the transverse photon, however,  an additional term appears in the cross section that is proportional to $m^2_f$.
Explicitly, the diffractive cross section for the transverse photon with quark mass reads
\beq
&&\frac{d\sigma^{\rm diff \; dijet}_{\rm T}}{dz_0dz_1d^2\p_0d^2\p_1}= \delta(z_0+z_1-1)2N_c\alpha_{em}(2\pi)^2\sum_fe^2_f
\int\frac{d^2\x_0}{(2\pi)^2} \int\frac{d^2\x_1}{(2\pi)^2} \int\frac{d^2\x_0'}{(2\pi)^2} \int\frac{d^2\x_1'}{(2\pi)^2} e^{-i\p_0\cdot\x_{00'}} e^{-i\p_1\cdot\x_{11'}}\nonumber\\
&&\times \;{\cal N}_{01}\; {\cal N}_{1^\prime 0^\prime}\Bigg\{ (z^2_0+z^2_1)\frac{\x_{01}\cdot\x_{0'1'}}{x^2_{01}x^2_{0'1'}}\left[\varepsilon_f x_{01} K_1(\varepsilon_f x_{01})\right] \left[\varepsilon_f x_{0'1'} K_1(\varepsilon_f x_{0'1'})\right] 
+m^2_f K_0(\varepsilon_f x_{01}) K_0(\varepsilon_f x_{0'1'})
\Bigg\}.
\eeq
By employing the same change of variables $(x_0, x_1, x_0', x_1')\mapsto (r, r', b, b')$, using full azimuthal symmetry, assuming that the dipole amplitude is independent of the relative angle between $\r$ and $\b$, the cross section for the transverse photon reads 
\beq
&&\frac{d\sigma^{\rm diff \; dijet}_{\rm T}}{dz_0dz_1d^2\p_0d^2\p_1}= \delta(z_0+z_1-1)\frac{2N_c\alpha_{em}}{(2\pi)^2}\sum_fe^2_f\nonumber\\
&&\times\; \Bigg\{(z^2_0+z^2_1)\bigg[\int_0^{+\infty}dr\int_0^{+\infty}db\; b\; J_0(b|\p_0+\p_1|)\;J_1\left( r \frac{|\p_0-\p_1|}{2}\right)\; \varepsilon_f r K_1(\varepsilon_f r){\cal N}(r,b)\bigg]^2\nonumber\\
&&\hspace{0.7cm}+m^2_f \; \bigg[\int_0^{+\infty}dr\int_0^{+\infty}db\; b\; J_0(b|\p_0+\p_1|)\;J_0\left( r \frac{|\p_0-\p_1|}{2}\right)\;K_0(\varepsilon_f r){\cal N}(r,b)\bigg]^2\Bigg\}.
\eeq
Summing up, the diffractive cross section with the inclusion of the mass reads
\beq
\frac{d\sigma^{\rm diff \; dijet}_{\rm T,L}}{dz_0dz_1d^2\p_0d^2\p_1}= \delta(z_0+z_1-1)\frac{2N_c\alpha_{em}}{(2\pi)^2}\sum_fe^2_f\; 
f_{\rm T,L}(z_0,z_1,\p_0,\p_1),
\label{xsectionwithmass}
\eeq 
with
\beq
f_{\rm L}(z_0,z_1,\p_0,\p_1)&=&4z_0z_1\varepsilon^2_f\bigg[ \int_0^{+\infty}dr\; r\int_0^{+\infty}db\; b\; {\cal N}(r,b) \; K_0(\varepsilon_f r)\; J_0(b|\p_0+\p_1|)\;J_0\left( r \frac{|\p_0-\p_1|}{2}\right) \bigg]^2,
\label{fL}
\\
f_{\rm T}(z_0,z_1,\p_0,\p_1)&=&(z^2_0+z^2_1)\bigg[ \int_0^{+\infty}dr\int_0^{+\infty}db\; b\; {\cal N}(r,b) \; \varepsilon_f r\;K_1(\varepsilon_f r) \;J_0(b|\p_0+\p_1|)\;J_1\left( r \frac{|\p_0-\p_1|}{2}\right) \bigg]^2\nonumber\\
&+&m^2_f\bigg[ \int_0^{+\infty}dr\; r\int_0^{+\infty}db\; b\; {\cal N}(r,b) \; K_0(\varepsilon_f r)\; J_0(b|\p_0+\p_1|)\;J_1\left( r \frac{|\p_0-\p_1|}{2}\right) \bigg]^2.
\label{fT}
\eeq

\subsection{Kinematics and final formulas}

The cross section can be written in the laboratory frame by introducing the limits of integrations on both $z_0$ and $z_1$ with the appropriate  $\theta$-functions:
\beq
\frac{d\sigma^{\rm diff \; dijet}_{\rm T,L}}{d^2\p_0d^2\p_1}&=&
\int_{-\infty}^{+\infty}dz_0\int_{-\infty}^{+\infty}dz_1 \, \theta\left(z_0-z_0^{\rm min}\right) \, \theta\left(z_1-z_1^{\rm min}\right) \, \theta\left(z_0^{\rm max}-z_0\right) \, \theta\left(z_1^{\rm max}-z_1\right) \, \delta\left(z_0+z_1-1\right) \nonumber\\
&&\hspace{3cm}\times \;  2N_c \frac{\alpha_{em}}{(2\pi)^2}\sum_{f} f_{\rm T,L}(z_0,z_1,\p_0,\p_1),
\eeq
with $f_{\rm T,L}(z_0,1-z_0,\p_0,\p_1)$  defined in Eqs. \eqref{fL} and \eqref{fT} and 
\beq
z_0^{\rm max} > z_0^{\rm min} > 0 \,  ,\\
z_1^{\rm max} > z_1^{\rm min} > 0 \, .
\eeq
The integration over $z_1$ can be performed trivially by realizing the $\delta$-function.
\beq
\frac{d\sigma^{\rm diff \; dijet}_{\rm T,L}}{d^2\p_0d^2\p_1}&=&
\int_{-\infty}^{+\infty}dz_0\,  \theta\left(z_0-z_0^{\rm min}\right) \, \theta\left(1-z_1^{\rm min}-z_0\right) \, \theta\left(z_0^{\rm max}-z_0\right) \, \theta\left(z_0-1+z_1^{\rm max}\right) \, 
 \nonumber\\
&&\hspace{1.5cm}\times \; 
 2N_c \frac{\alpha_{em}}{(2\pi)^2}\sum_{f} f_{\rm T,L}(z_0,1-z_0,\p_0,\p_1).
\label{xsection_z0}
\eeq
Eq. \eqref{xsection_z0} has a non-zero result only if 
\beq
1-z_1^{\rm min} > z_0^{\rm min}\, , \\
z_0^{\rm max} > 1-z_1^{\rm max}.
\eeq 
In the laboratory frame $z_{0,1}$ reads
\beq
z_{0,1}= \frac{M_T\sqrt{\p^2_{0,1}+m_f^2}}{2q^+P^-}\exp\left[ \eta^{lab}_{0,1}+
\operatorname{arcosh}\left(\frac{E_p}{M_T}\right)
\right],
\label{z01Lab}
\eeq
where $M_T$ is the mass of target and $\eta^{lab}_{0}, \eta^{lab}_{1}$ denote the rapidities of the jets in the laboratory frame. We denote the proton beam energy and the centre-of-mass energy of the photon-proton system by $E_p$ and $W_{\gamma p}$, respectively. Assuming that $E_p\gg M_T$, $m_f\ll |\p_{0,1}|$ and $2P^-q^+\simeq W_{\gamma p}^2$, Eq. \eqref{z01Lab} can be approximated as 
\beq
z_{0,1}\simeq \frac{2E_p|\p_{0,1}|}{W_{\gamma p}^2}\, e^{\eta_{0,1}^{lab}}\, ,
\eeq
where one can read off the minimum and maximum values of $z_{0,1}$:
\beq
z_{0,1}^{\rm min, max}\simeq \frac{2E_p|\p_{0,1}|}{W_{\gamma p}^2}\, e^{{\eta_{0,1}^{lab\  {\rm min, max}}}}.
\eeq

Note that, in principle,  functions   $f_{\rm T,L}(z_0,1-z_0,\p_0,\p_1)$ also depends on $x_g$ which enters in the dipole amplitude (due to higher order corrections). The light-cone variable $x_g$ corresponds to Bjorken-$x$ in inclusive DIS. The explicit expression for $x_g$ reads
\beq \label{xg}
x_g=\frac{\p_1^2}{W_{\gamma p}^2}\left( \frac{z_0}{1-z_0}\right)+\frac{\p_0^2}{W_{\gamma p}^2}\left( \frac{1-z_0}{z_0}\right)-\frac{\p_0\cdot\p_1}{W_{\gamma p}^2}+\frac{Q^2}{W_{\gamma p}^2}\ ,
\eeq
where $W_{\gamma p}^2=(P+q)^2=M_T^2-Q^2+2P\cdot q$.  The expression above is obtained via energy-momentum conservation.

Finally, note that we have obtained so far the diffractive dijet cross section  $\sigma^{\gamma^*p}$ in $\gamma^{(*)}$-proton collisions. One can relate $\sigma^{\gamma^*p}$ to the diffractive dijet cross section $\sigma^{ep}$ in electron-proton (ep) collisions in DIS via
\beq
\frac{d\sigma^{ep}}{dxdQ^2}=\frac{\alpha_{em}}{\pi x Q^2}\left[ \left( 1-y+\frac{y^2}{2}\right) \; \sigma^{\gamma^*p}_{\rm T}+(1-y)\sigma^{\gamma^*p}_{\rm L}\right], 
\eeq
where $y=Q^2/(sx)$ is the inelasticity variable and $\sqrt{s}$
denotes the centre-of-mass energy of the ep collision. In this
 expression, we neglect the contribution of the $Z$ boson that becomes sizeable only at very large $Q^2$.

\begin{figure}[t]       
\includegraphics[width=0.45\textwidth,clip]{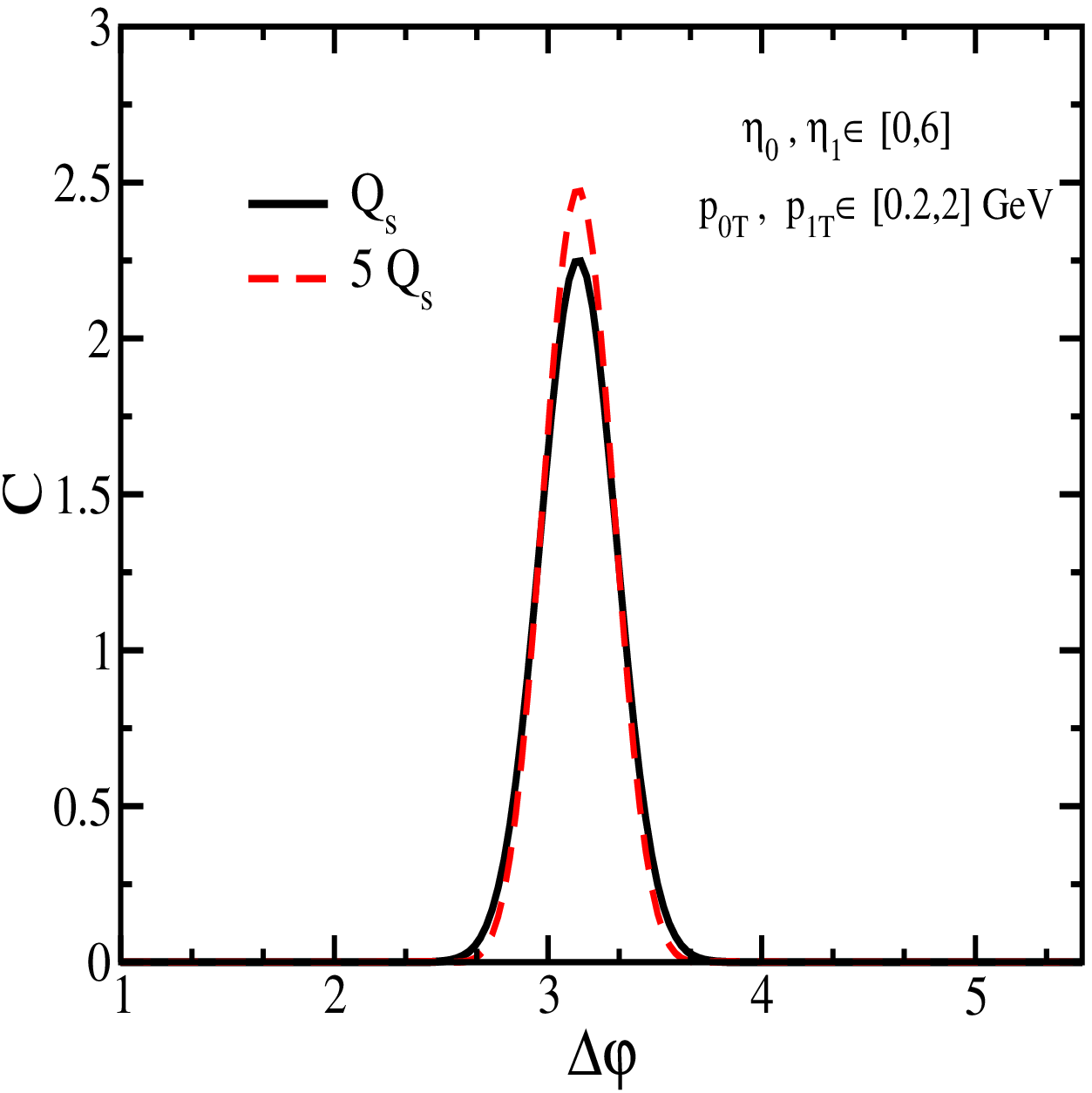}\hskip 1cm\includegraphics[width=0.45\textwidth,clip]{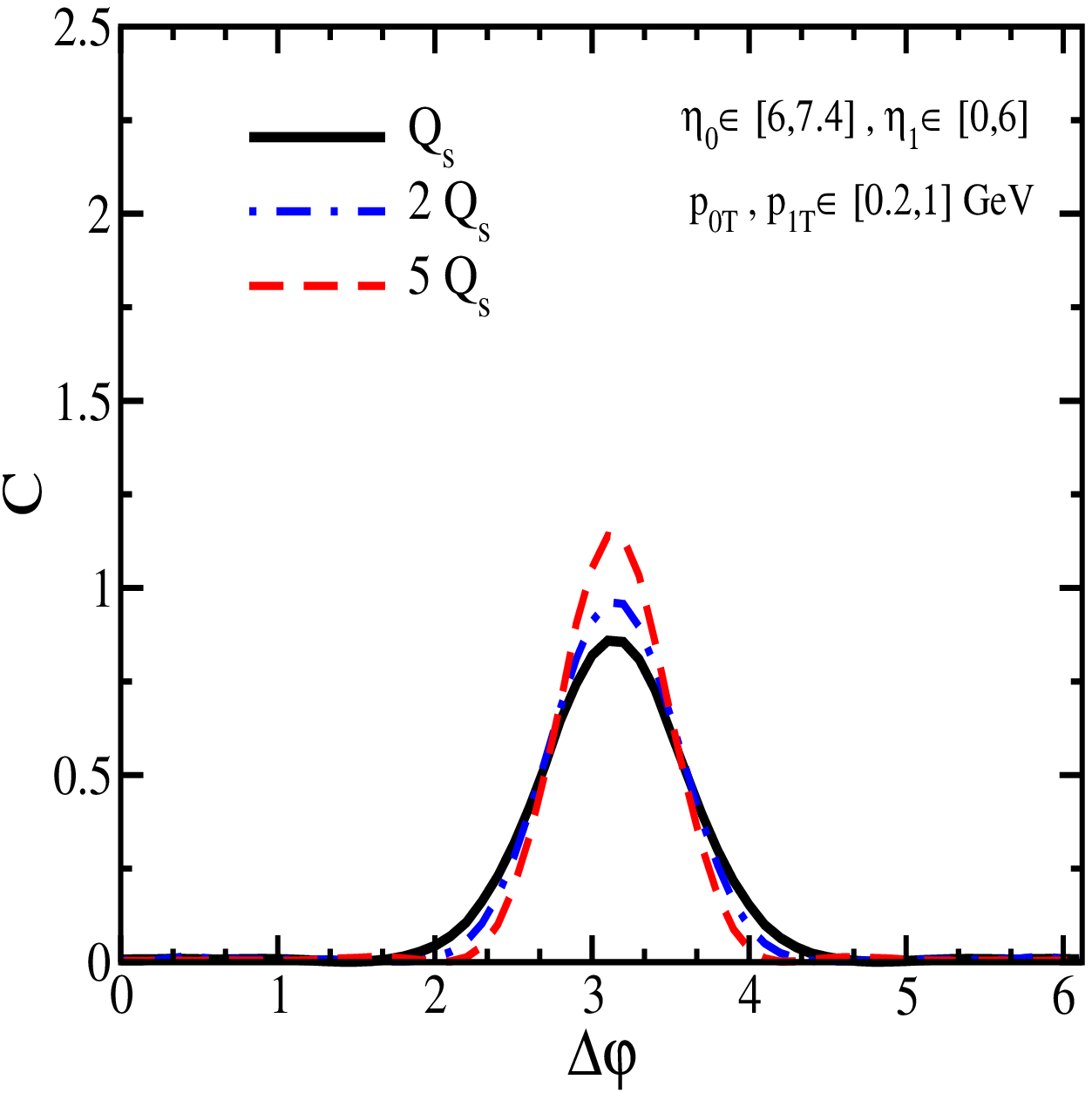}
\caption{Diffractive dijet photo-production correlations $C$ as a function of the angle $\Delta\varphi$ between the two jets, in the IP-Sat model with different saturation scales $Q_s$, $2Q_s$ and $5Q_s$ ($Q_s$ denotes the saturation scale of  a proton that enters the dipole scattering amplitude), for two kinematic bins in rapidities $\eta_0, \eta_1$ and transverse momenta $p_{0T}$, $p_{1T}$ of the two jets given in the plot.  We consider $E_p=5$ TeV and  $W_{\gamma p}=4$ TeV for all curves. }
\label{f-angle1}
\end{figure}

\section{Numerical results}
We first focus on azimuthal angle correlations of diffractive dijet production in $\gamma^{(\star)}$-p(A) collisions.  We define the azimuthal correlation in the following form \cite{amir-photon,2gamma}:
\begin{equation}\label{az}
C(\Delta \varphi)=\frac{d\sigma^{\gamma^{\star} p \rightarrow q\bar{q} p}} {d\p_{0}\, \p_{1} d\Delta\varphi}{\Bigg /} \int_{0}^{2\pi} d\Delta\varphi \frac{d\sigma^{\gamma^{\star} p \rightarrow q\bar{q} p}} {d\p_{0}\, \p_{1} d\Delta\varphi}\ , 
\end{equation}
where the angle $\Delta \varphi$ 
is the difference between the azimuthal angles of the two jets. The function $C(\Delta \varphi)$ has the meaning of the probability of  semi-inclusive diffractive dijet pair production at a certain kinematics and angle  $\Delta \varphi$, triggering the same production with the same kinematics integrated over $\Delta \varphi$.

The main ingredient of the cross section for diffractive dijet DIS in \eq{xsection_z0} is the universal $q\bar{q}$ dipole-target amplitude $\mathcal{N}(r,b,x)$. It is universal since the same dipole amplitude in the fundamental representation also appears in the cross sections for structure functions in DIS,  exclusive diffractive vector meson production and deeply virtual Compton Scattering (DVCS). As seen in \eq{xsection_z0}, the impact-parameter dependence of the
dipole amplitude is crucial for describing exclusive diffractive dijet process.  We use two well-known impact-parameter dependent saturation models, the so-called IP-Sat \cite{ipsat0,ipsat} and the b-CGC \cite{bcgc0,bcgc} models which both have been very successful in phenomenological applications from HERA to the Relativistic Heavy Ion Collider (RHIC) and the LHC, see for example \cite{pp-amir,pp-raju}. The parameters of the IP-Sat and the b-CGC models were determined via a fit to the recently released high-precision combined HERA data. These two models  provide an equally good description of all available small-$x$ HERA data for diffractive vector meson and DVCS production \cite{ipsat,bcgc,na}. The saturation scale in the IP-Sat and the b-CGC models depends on the impact-parameter and the light-cone parameter $x_g$.   Note that the light-cone parameter $x_g$ depends on kinematics, see \eq{xg}.  For a typical small $x_g$ here, the saturation scale can change from a few MeV to a few GeV depending on the impact parameter, see Fig. 3 in Ref.\,\cite{bcgc}. In order to compare with non-saturation models, we employ the-so-called 1-Pomeron model \cite{na}  that is the leading-order pQCD expansion for the dipole amplitude in the color transparency region, as opposed to the saturation regime.  Note that the dipole amplitude $\mathcal{N}(r,b,x)$ is the only external input here, with its free parameters already fixed via a fit to other reactions. Therefore our results here can be considered as free-parameter calculations. 

In the following, we consider photo-production ($Q\approx 0$) of light quark jets at a proton beam energy $E_p=5$ TeV, and a centre-of-mass energy of the photon-proton system $W_{\gamma p}=4$ TeV.  The assumed values for $E_p$ and  $W_{\gamma p}$ can be taken as representative of the energies achievable at the LHC or in future ep colliders. The main features of our results will be unchanged by using different values for these quantities.  We also show the results as a function of $W_{\gamma p}$. 
For the numerical computation we focus  on low transverse momenta of the produced jet pairs. Note that this kinematics is mostly relevant for probing saturation effects, as we will demonstrate here. Note also that the free parameters of the impact-parameter dependent saturation model employed here, were obtained via a fit to small-$x$ ($x<0.01$)  data at HERA corresponding to low virtualities $Q^2<40\div 45\,\text{GeV}^2$ \cite{ipsat,bcgc}. Therefore these models should  be considered less reliable for large  $p_T>6$ GeV.  

For numerical purposes,  for the case of photo-production we take the virtuality $Q^2\approx 0.04 \div 0.1 \,\text{GeV}^2$. Note that at HERA, photon virtualities close to zero are  considered as photo-production, see for example Refs.\,\cite{q0}. We use the same convention. Therefore, our results should be considered as an extrapolation to the case of $Q=0$. Such an extrapolation is reasonable and is consistent with existing HERA data for $Q^2 \le 0.5\,\text{GeV}^2$ \cite{ipsat0,ipsat,bcgc0,bcgc}. Moreover, for the light quark mass we have taken $m_q\sim 4\div 10 $ MeV. Note that the light quark masses were determined via a fit to the recent combined HERA data, and it was shown in Refs.\,\cite{ipsat,bcgc} that light quark masses $m_q$ about the current quark mass are consistent with the recent combined HERA data at small-$x$. Therefore, the integrals in the case of the 1-pomeron model are convergent.

\begin{figure}[t]       
 \includegraphics[width=0.45\textwidth,clip]{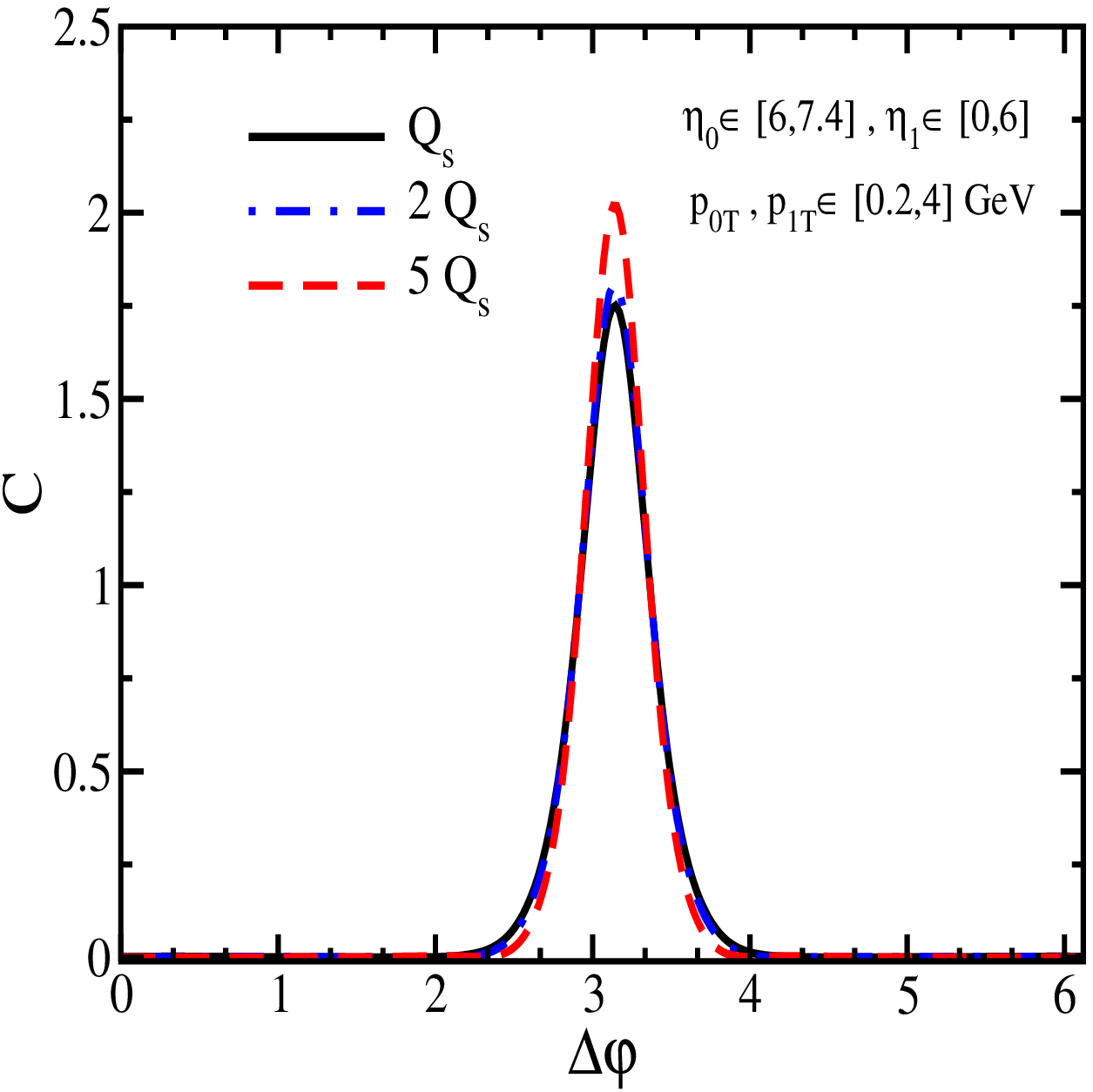} \hskip 1cm \includegraphics[width=0.45\textwidth,clip]{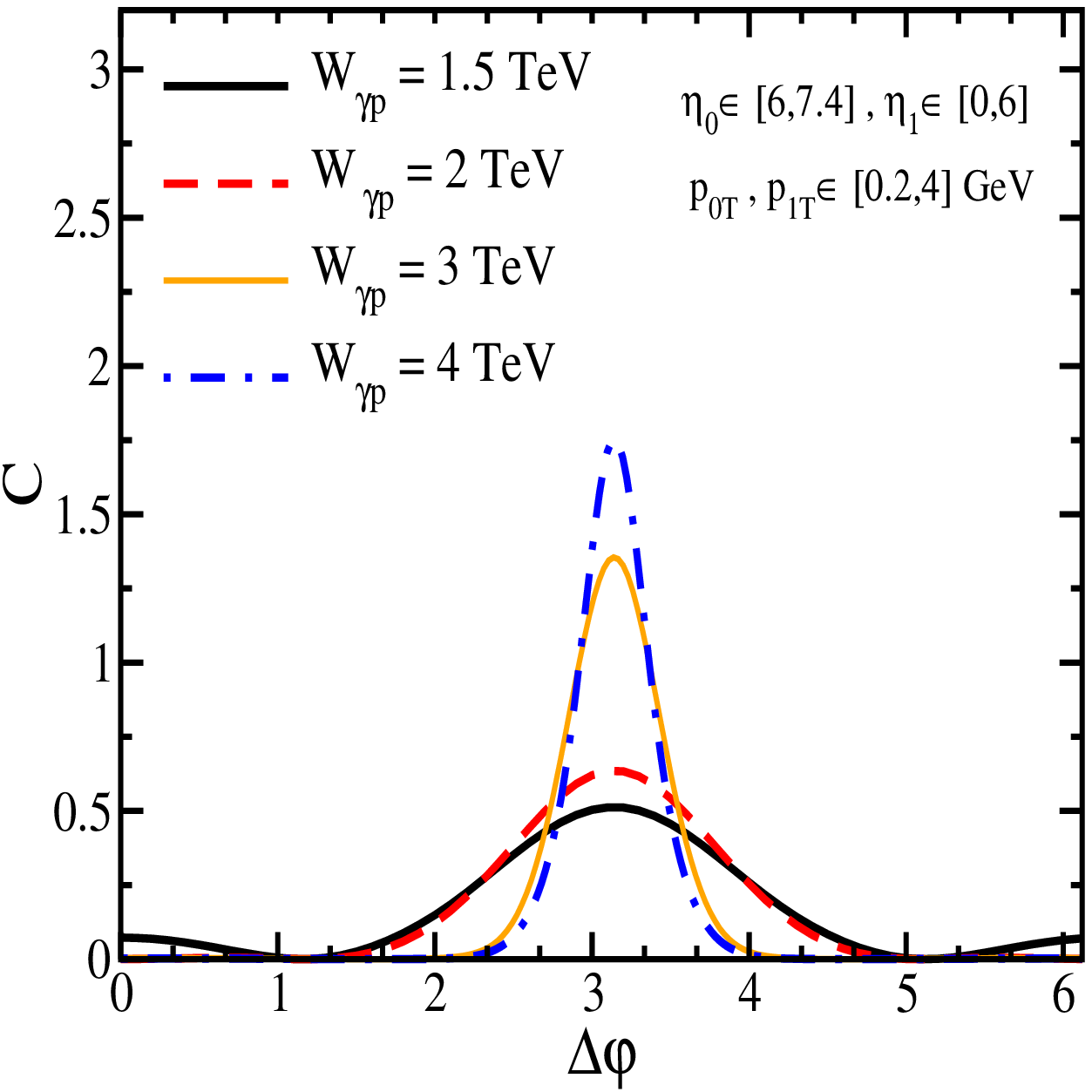}
\caption{Left: saturation scale dependence of diffractive dijet photo-production correlations $C$ as a function of the angle $\Delta\varphi$ between the two jets, in the IP-Sat model with different saturation scales $Q_s$, $2Q_s$ and $5Q_s$ ($Q_s$ denotes the saturation scale of  a proton that enters the dipole scattering amplitude) at a fixed $W_{\gamma p}=4$ TeV.   Right: energy dependence of diffractive dijet photo-production correlations $C$ as a function of the angle $\Delta\varphi$ between the two jets at various center-of-mass energy of the photon-proton system $W_{\gamma p}$. All curves in left panels are obtained by using  the IP-Sat model for a proton target with the saturation scale $Q_s$. We consider $E_p=5$ TeV in both panels. }
\label{f-angle1-n}
\end{figure}

\begin{figure}[t]       
\includegraphics[width=0.45\textwidth,clip]{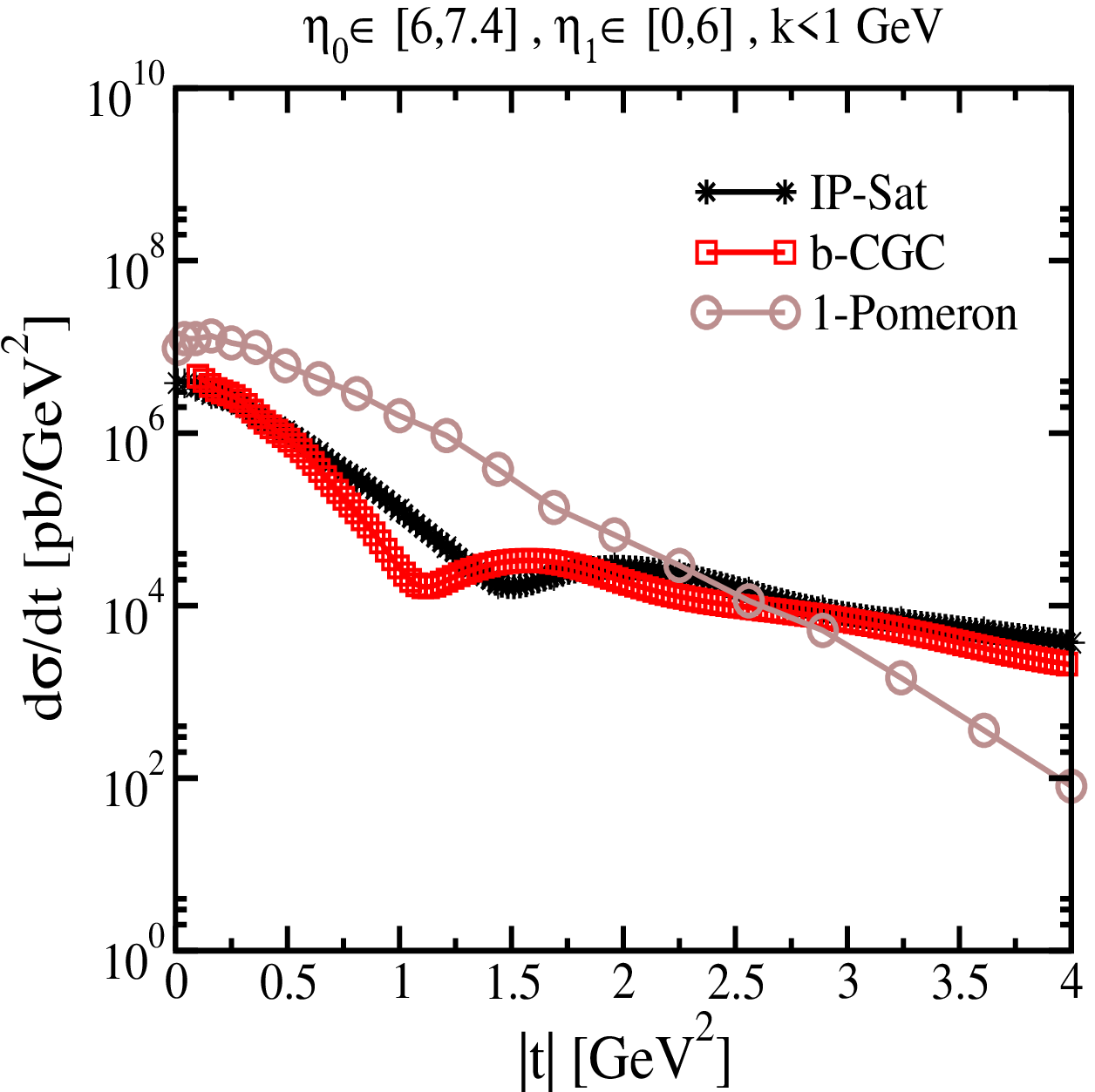}\hskip 1cm\includegraphics[width=0.45\textwidth,clip]{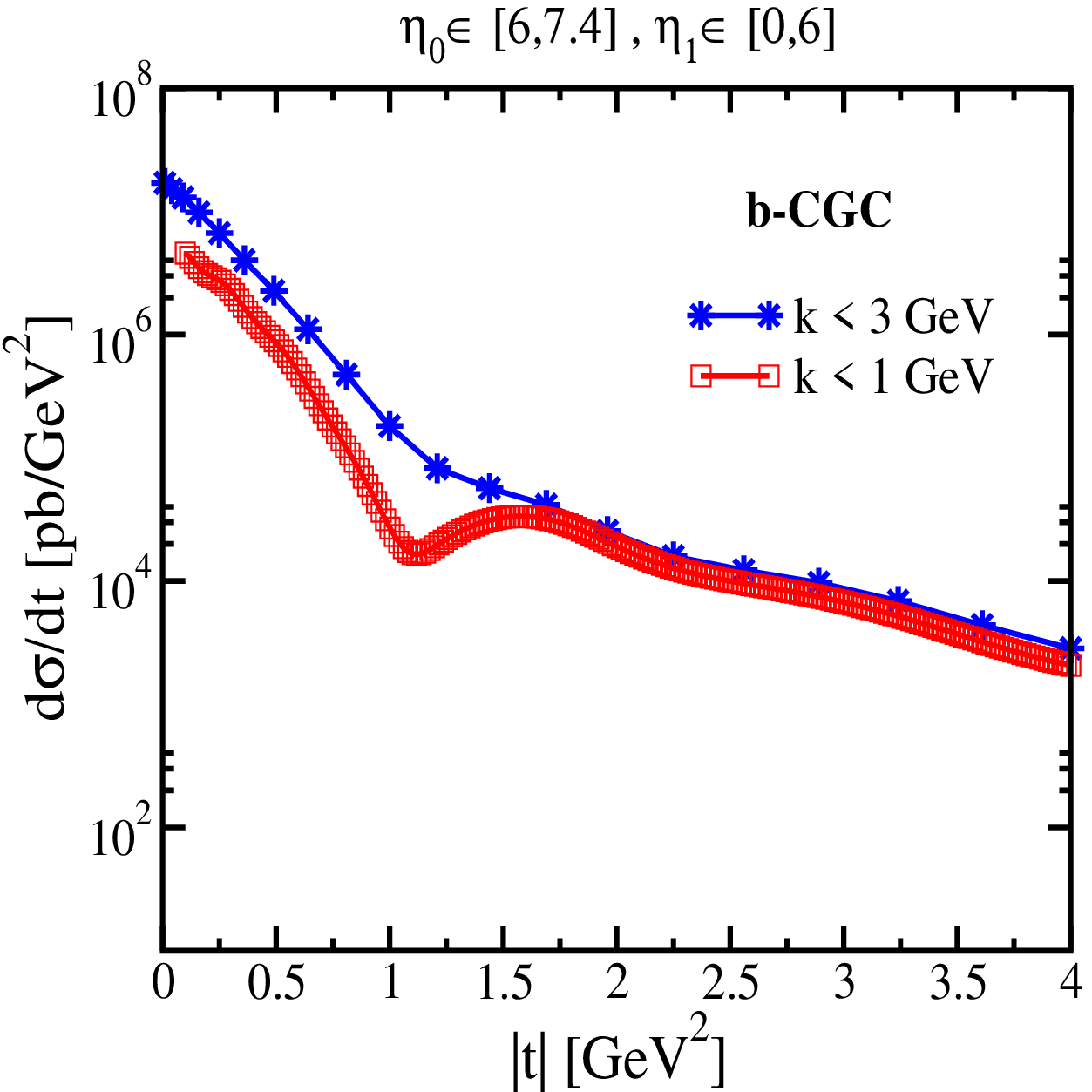}
\caption{Differential dijet photo-production cross sections as a function of $|t|$ for the IP-Sat, the b-CGC and the non-saturation 1-Pomeron models in $\gamma^{(\star)}$-p(A) collisions. Left: The differential cross sections are integrated over $\k<1$ GeV. Right:  The differential cross sections are compared in the b-CGC model for two different bins in $\k$. In both panels, we consider $E_p=5$ TeV, $W_{\gamma p}=4$ TeV, and we integrate over the rapidities of two jets $\eta_0$ and $\eta_1$ within $[6,7.4]$ and $[0,6]$, respectively. }
 \label{f-t1}
\end{figure}

In \fig{f-angle1}, we show  diffractive dijet correlations $C(\Delta \varphi)$ defined via \eq{az} as a function of the angle between the two jets $\Delta \varphi$, in $\gamma^{(\star)}$-p(A) collisions. In the first case, shown in \fig{f-angle1} left, we integrate over the rapidities of two jets around mid-rapidity $\eta_0, \eta_1\in [0,6]$ and  also  over the transverse momenta of the two jets within  $0.2\leq \p_{0}, \p_{1} [\text{GeV}] \leq 2$. Note that all rapidities in this section are given in the laboratory frame. 
In the other case shown in \fig{f-angle1} right, we consider the two jets to be in two different rapidities intervals, one at forward and the other at mid rapidity, performing the integral over the rapidities and transverse momenta\footnote{ In the plots transverse momenta are denoted by subscript T.} of two jets within  $\eta_0\in [7.4, 6]$, $\eta_1\in [0,6]$, and $0.2\leq \p_{0}, \p_{1} [\text{GeV}] \leq1$. In order to further show the sensitivity of the diffractive dijet correlations to saturation physics, we show in \fig{f-angle1} our results obtained in the IP-Sat model with different saturation scales, one that corresponds to the saturation scale of a proton $Q_s$ extracted from a fit to the HERA data and one with an enhanced saturation scale $2Q_s$ which roughly simulates its typical magnitude for a heavy nucleus in {\it minimum-bias} collisions \cite{ja,pa-amir,pa-jav}. We also obtained the results assuming an enhanced saturation scale  $5Q_s$ which effectively considers events with high-multiplicity \cite{high-mu1,high-mu2}. It can be seen that the correlations are suppressed and broadened in the case where one jet is  at very forward rapidity compared to the case where both jets are around mid-rapidities.  It is also seen in \fig{f-angle1} that the diffractive dijet back-to-back correlations are relatively enhanced in a model with a larger saturation scale. This feature is characteristic of the diffractive process and opposite to  inclusive production \cite{amir-photon,2gamma,2h-jav}. 
A possible explanation can be that, in order to keep the color neutrality of the dijet system, required by its diffractive nature, the production becomes dominated by $q \bar q$ pairs of smaller transverse size with increasing saturation momentum. These pairs of smaller transverse size correspond to larger relative transverse momentum between the jets, and thus their angular correlation is  less affected by saturation effects. Besides, as we discuss below, production is shifted to larger impact parameters where saturation effects are smaller\footnote{Note also that in contrast to inclusive production where the back-to-back correlation can be unbalanced by the existence of an extra scale in the system like the saturation scale, in the case of diffractive production, in principle, the extra scale in the system can partially get balanced by a momentum transfer to the target.}. 

In \fig{f-angle1-n} left panel, we show the saturation scale dependence of the correlation $C$ for transverse momentum of jets within $p_{0T},p_{1T}\in [0.2, 4]$ GeV. Comparing \fig{f-angle1-n} (left panel) and \fig{f-angle1} (right panel), it is seen that integrating over larger transverse momenta, makes the correlation $C$ be less sensitive to the saturation effect as one may expect. 
 Note also that the relation between $x_g$, $p_{0T},p_{1T}$ and $W_{\gamma p}$ is non-trivial, see \eq{xg}. Therefore, a priori it is not clear which kinematic variable can be a better probe of the saturation effect here. In \fig{f-angle1-n} right panel, we show the energy $W_{\gamma p}$ dependence of the correlation $C$ for transverse momentum of jets within $p_{0T},p_{1T}\in [0.2, 4]$ GeV.  It is seen from this plot that the scanning the correlation function $C$ at different centre-of-mass energy of the photon-proton system $W_{\gamma p}$,  can be a good probe of the saturation effect, and it leads to a sizeable effect even at rather large transverse momenta of the produced jets (integrating over transverse momenta of jets in $p_T<4$ GeV). 

\begin{figure}[t]       
\includegraphics[width=0.45\textwidth,clip]{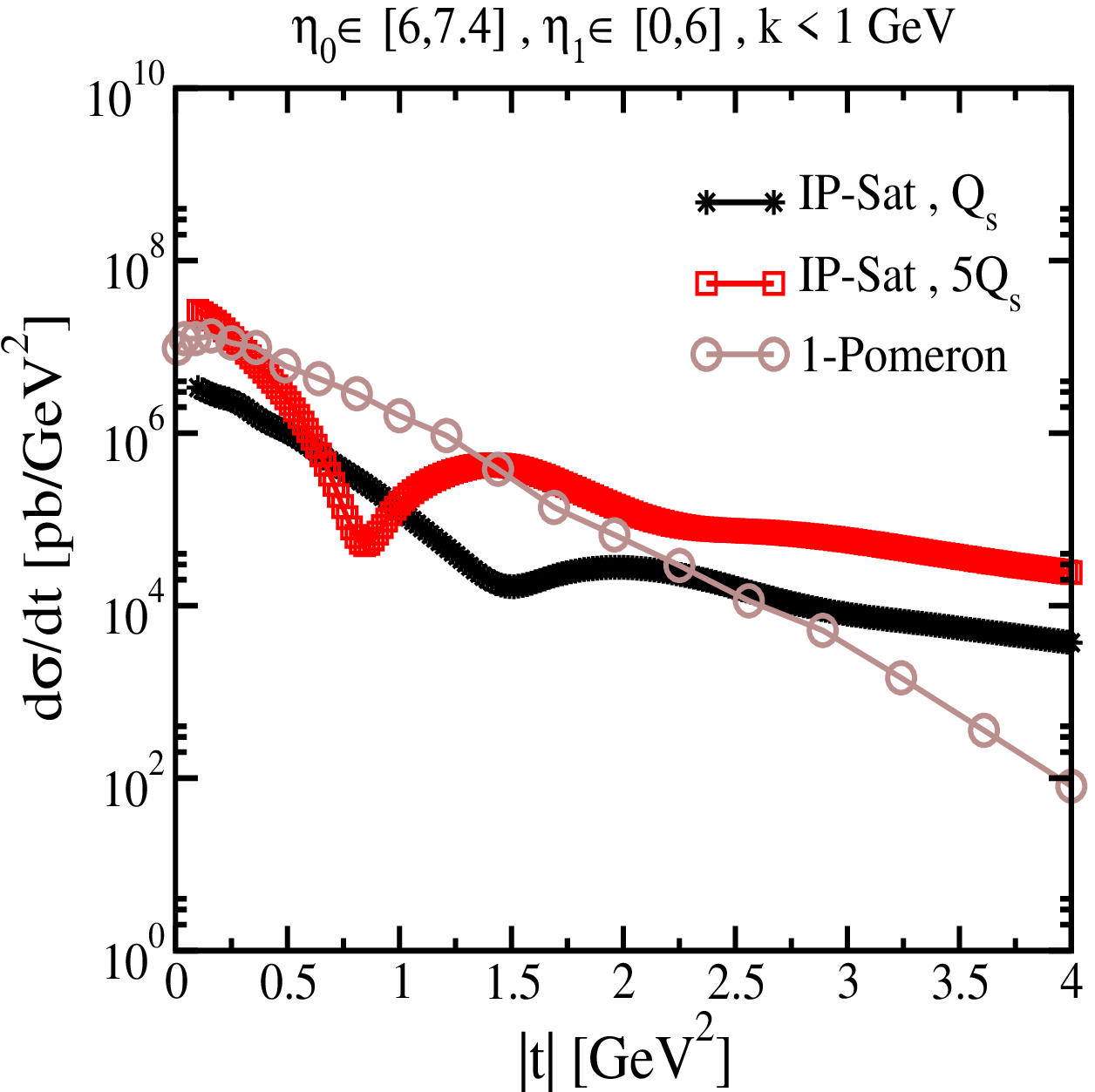}\hskip 1cm\includegraphics[width=0.45\textwidth,clip]{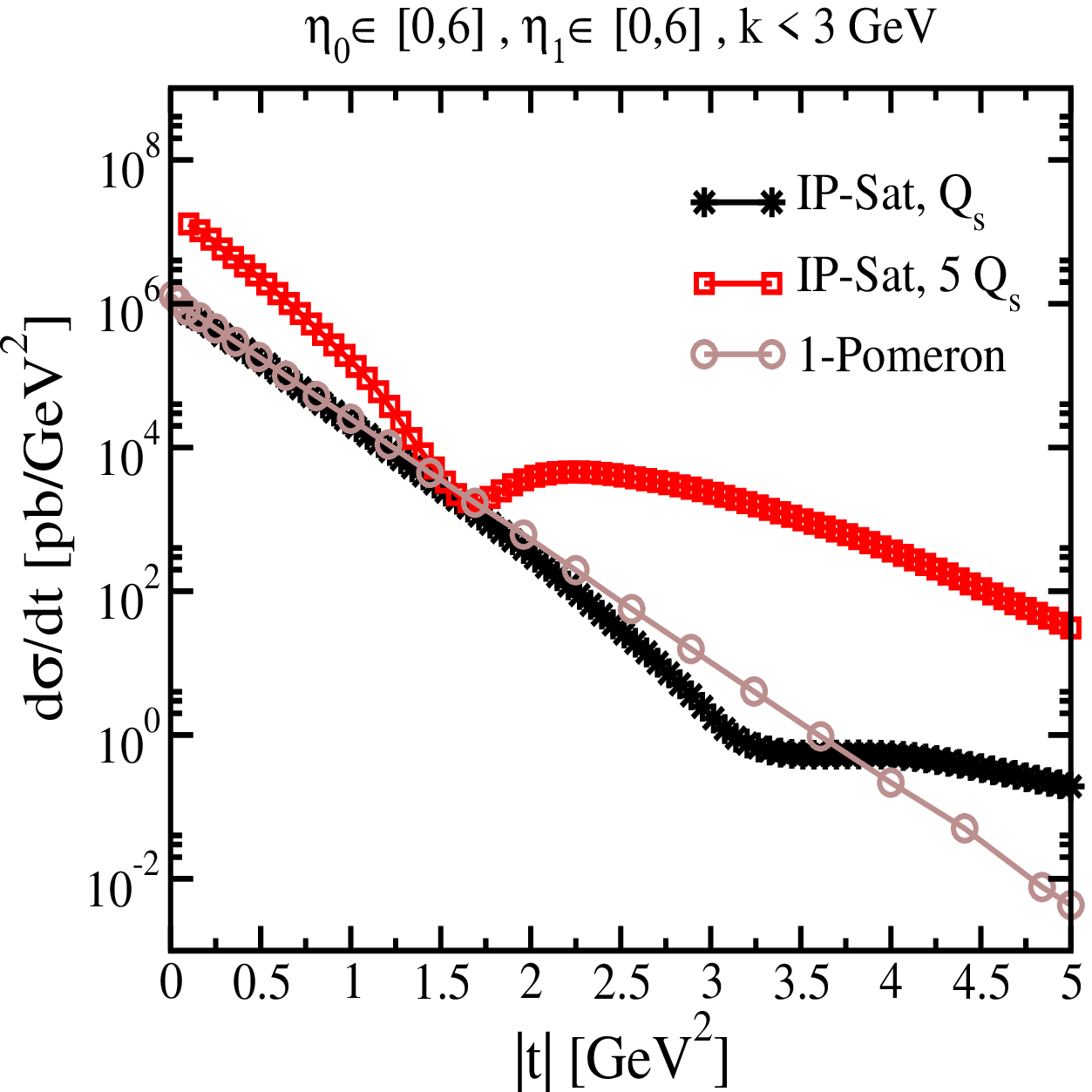}
\caption{Differential dijet photo-production cross sections as a function of $|t|$ for the IP-Sat (with different saturation scale $Q_s$) and the non-saturation 1-Pomeron models in $\gamma^{(\star)}$-p(A) collisions.  In the left panel we integrate over $k<1$ GeV and  the rapidities of the two jets within  $\eta_0\in [6,7.4]$ and $\eta_1\in [0,6]$, while in the right panel we integrate over $k<3$ GeV and the rapidities of both jets within $[0,6]$. We consider $E_p=5$ TeV and $W_{\gamma p}=4$ TeV for all curves.}
 \label{f-t2}
\end{figure}     

In order to further investigate if the enhancement of the back-to-back correlations shown in \fig{f-angle1} is truly due to  saturation physics, we study the $t$-distribution of diffractive dijet production. To this end,  we change  variables ($\p_0$,  $\p_{1}$) to ($\Delta=\p_0+\p_1$,  $\k=\frac{1}{2}(\p_0-\p_1)$) in the cross section, \eq{mless}. Then we have $|t|=\Delta^2$ with  $t$ being the squared momentum transfer.  
Note that the cross sections of diffractive dijet production and diffractive vector meson production behave quite similarly in the small-$x$ region: In both cases,  the cross section can be written in terms of the dipole amplitude, and in both cases  $\Delta$ and $\b$ are Fourier conjugates, $b \sim 1/\sqrt{|t|}$. In Ref.\,\cite{na}, it was shown that the $t$-distribution of diffractive vector meson production exhibits a dip-like structure in the saturation regime due to the unitarity features of the color dipole amplitude. Therefore, one may expect that if the enhancement of the back-to-back correlations  shown in \fig{f-angle1}  is due to the saturation physics, one should also see a dip-like structure for the $t$-distribution of diffractive dijet production in the saturation region (in the same kinematic region considered in \fig{f-angle1}).

In \fig{f-t1} left , we show the $t$-distribution of diffractive dijet in the same kinematics region considered in \fig{f-angle1} (with one of the two jets at forward rapidity). It is seen that in the saturation models (IP-Sat, and  b-CGC), the diffractive dijet $t$-distribution has a dip while in the non-saturation model (1-Pomeron), there is no dip-like structure. On the right panel of \fig{f-t1}, it is shown that the dip disappears by increasing the value of $k$. This is due to the fact that lowering $k$, enhances the saturation effects and consequently the dip in the $t$-distribution becomes more pronounced.   In order to further show that the saturation dynamics is main cause of the dip-structure, in \fig{f-t2} we show the effect of changing the saturation scale in the $t$-distribution of diffractive dijet production in two kinematic regions with $k<3$ GeV and $k<1$ GeV. It is seen that the dip in the $t$-distribution moves toward lower $t$ and becomes stronger by increasing the saturation scale, and disappears in the non-saturation 1-Pomeron model. 

Note that although dips are due to the shape in impact parameter that has a non-perturbative origin, the main difference between a dipole model with linear and non-linear evolution (i.e, incorporating saturation effects through some specific model as those employed in this work), is that the former does not lead to the black-disc limit and, therefore, the dips do not systematically shift toward lower $|t|$ by increasing saturation scale, while the latter does. Non-linear evolution evolves any realistic impact-parameter profile, like a Gaussian or Woods-Saxon distribution, and makes it closer to a step-like function in impact parameter by limiting the growth in the denser centre. This leads to the appearance of dips with non-linear evolution even if the dips were not present at the initial condition at low energies or for large $x$, or to the receding of dips towards lower values of $|t|$ even if they were already present in the initial condition. Note that small $|t|$ corresponds to  peripheral collisions when saturation physics should be less important. The main features of the $t$-distribution of diffractive dijet production are similar to those seen in diffractive vector meson production \cite{na}. The enhanced back-to-back correlation with increasing saturation scale and emergence of dip in $t$-distribution appears to be a universal feature of diffractive dijet production, irrespective of the mechanism by which the saturation scale is increased.

Finally we have made some studies of the effect of color dipole orientation or the existence of a correlation between $\r$ and $\b$ in the master \eq{azy} for diffractive dijet production in DIS. Our results, with the aim of not performing a realistic calculation but of showing that indeed an observable effect can be obtained, are shown in the Appendix.

\section{Conclusions}
\label{conclu}
In this paper we analyse exclusive  dijet production in coherent diffractive processes in DIS and real (and virtual) photon-hadron collisions in the CGC framework at leading order. In contrast to  inclusive dijet production in DIS where the cross section depends on both two- and four-point Wilson line correlations (thereby involving the Weizs\"acker-Williams gluon distribution), diffractive dijet production in DIS only depends on the two-point function (thus only on the dipole gluon distribution), see \eq{xsection_z0}. This may be considered as an advantage of diffractive over inclusive dijet production, since the dipole amplitude, as a solution of the BK evolution equation, has been well constrained by experimental data in different reactions. In contrast, the four-point function is less theoretically known, and remains rather unconstrained by current experimental data. 
We also show that the diffractive dijet cross section is sensitive to the color-dipole orientation in the transverse plane, a new ingredient which has not been previously explored, see  Eqs.\,(\ref{Lf1},\ref{Tf1}). This  feature of  diffractive dijet production may provide complementary information about possible correlations between the $q\bar{q}$ dipole transverse separation $\r$ and the dipole impact parameter $\b$, see the Appendix. Note that the impact-parameter profile of the dipole amplitude entails intrinsically non-perturbative physics, and a full description of the impact-parameter profile of the collisions seems to be beyond the QCD weak-coupling approach to small-$x$ physics. Nevertheless, there has recently been several attempts to extract information about the impact-parameter $\b$ dependence of the dipole amplitude via the BK evolution equation. These studies indeed indicate non-trivial correlations between $\r$ and $\b$ \cite{b-bk}. A detailed phenomenological study of the color dipole orientation in diffractive dijet production is beyond the scope of the current paper. It was shown that correlations between $\r$ and $\b$ in dipole amplitude, can naturally cause azimuthal asymmetry in hadronic collisions \cite{v2-dipole}. Nevertheless, most of the models for the dipole amplitude in the literature do not include this feature. Hence, we provide corresponding simplified expressions for the numerical analysis.

We also investigate diffractive dijet correlations and $t$-distributions in $\gamma^{(*)}$-h collisions and analyse their sensitivity to gluon saturation effects in the small-$x$ kinematic region. We show that an increase of the saturation scale produces an enhancement of away-side correlations. This feature seems to be unique to  diffractive production, and is in drastic contrast to the inclusive two-particle correlations such as inclusive dijet \cite{2h-jav}, inclusive photon-hadron \cite{amir-photon}, and inclusive di-photon \cite{2gamma} production in the CGC framework. We also find that the $t$-differential cross section of diffractive dijet production in high-energy collisions offers a unique opportunity to probe the saturation regime and discriminate among models. It exhibits a dip-type structure in  saturation models while, in a non-saturation model, dips are either absent or expected to lie at larger $|t|$ and not to shift towards smaller values of  $|t|$ with increasing  saturation scale (or decreasing $x$). This behavior  is very similar to the one found in diffractive vector meson production \cite{na}, indicating the universality of the underlying dynamics  due to the initial-state effects.

In order to understand more rigorously the implications of  gluon saturation in the
proton wave function on diffractive processes, it is indispensable to systematically investigate the effect
of higher order contributions beyond the current leading-order approximation, see also Refs.\, \cite{Mueller:2012uf,Mueller:2013wwa,dijet2,Balitsky:2010ze,Balitsky:2012bs,Beuf:2011xd}. 
Nevertheless, the requirement of color neutrality in diffractive processes should pose significant constraints on higher order corrections\footnote{In Refs.\,\cite{Mueller:2012uf,Mueller:2013wwa} it was shown that  inclusive dijet production at leading order is subject to large corrections due to  the Sudakov double-logarithm resummation. Diffractive production was not considered in those references but, following the same line of arguments outlined there, color neutrality seems to suggest that 
double logarithmic corrections should not be present in diffractive dijet DIS. However, the Sudakov single logarithms may be non-vanishing. This effect remains to be studied.}. A detailed study of diffractive dijet production at full one-loop level is postponed to a future publication. Also, an analysis of jet reconstruction in more realistic experimental environments would be required to establish the practical feasibility of this kind of studies.

\begin{figure}[t]       
\includegraphics[width=0.45\textwidth,clip]{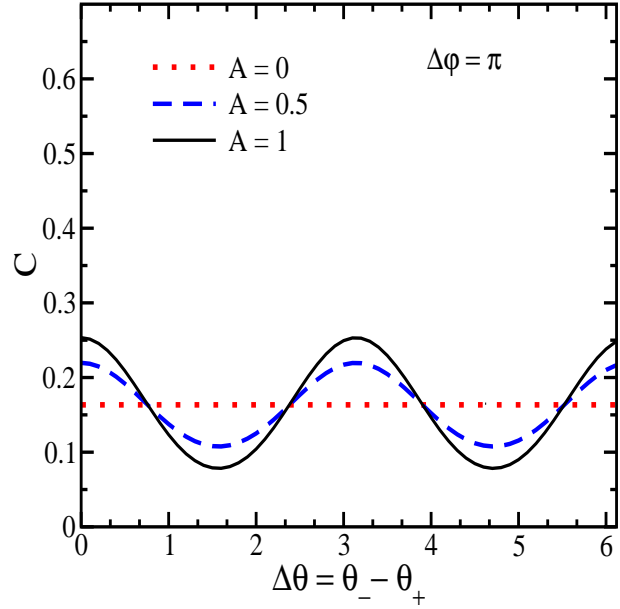}
\caption{Diffractive dijet photo-production correlations $C$ as a function of the angle $\Delta\theta=\theta_{-}-\theta_{+}$ between two vectors $\Delta=\p_0+\p_1$ and  $\k=\frac{1}{2}(\p_0-\p_1)$, at a fixed $\Delta\varphi=\pi$ (back-to-back jets) for various value of $A$ which determines the polarization of the dipole amplitude.  We integrate over the rapidities of two jets $\eta_0, \eta_1\in [0,6]$, $\eta_0, \eta_1\in [6,7.4]$  and  also  over the transverse momenta of the two jets within  $0.2\leq \p_{0}, \p_{1} [\text{GeV}] \leq 2$. We consider $E_p=5$ TeV and $W_{\gamma p}=4$ TeV for all curves.}
 \label{f-teta}
\end{figure}

\appendix
\label{app}
\section{}
In this Appendix we consider the effect of color dipole orientation or the existence of a correlation between $\r$ and $\b$ in the master \eq{azy} for diffractive dijet production in DIS. Our aim is not performing a realistic calculation but showing that indeed an observable effect can be obtained. To this end, we assume that the dipole amplitude has the following  form: 
\begin{equation}
{\cal N}(\r,\b)={\cal N}(r,b,\theta_r-\theta_b)=1-e^{-\frac{Q_s^2(b)}{4} r^2\left(1+A\cos^2(\theta_r-\theta_b)\right)}\, ,
\end{equation}
where  $\theta_r, \theta_b$ are the angles of vectors $\r$, $\b$ with respect to a reference vector, respectively. Parameter $A$ controls the size of the correlation, namely a dipole amplitude without correlation corresponds to $A=0$ that is the case for the IP-Sat and the b-CGC  models. In order to get an expression as simple as possible, we assume  $Q_s^2r^2A/4\ll 1$,  neglect terms suppressed by powers of $k r$ or $b \Delta$, and re-exponentiate the final result\footnote{No firm justification exists for this procedure and, therefore,  the final result can be considered simply as an ad hoc model.}. Proceeding in this way we get  
\beq
\int \frac{d^2\r}{(2\pi)^2}\int \frac{d^2\b}{(2\pi)^2} e^{-i\b\cdot(\p_0+\p_1)}e^{-i\r\cdot(\p_0-\p_1)/2}{\cal N}(\r,\b)K_0(\varepsilon|\r|)&\simeq&
\int_0^{+\infty}\frac{dr}{2\pi}r\int_0^{+\infty}\frac{db}{2\pi}b \;J_0\left(b|\p_0+\p_1|\right)\nonumber\\
&\times& \; J_0\left(r\frac{|\p_0-\p_1|}{2}\right){\cal N}(r,b, \theta_{-}-\theta_{+})K_0(\varepsilon r),
\eeq
where $\theta_+$ and $\theta_-$ denote the angles of vectors  $(\p_0+\p_1)$ and $(\p_0-\p_1)$ with respect to a reference vector, respectively. \eq{azy2} can be treated in an analogous manner. Therefore, a possible azimuthal correlation between $\vec r$ and $\vec b$ in the dipole amplitude leads to an observable effect of non-zero correlation between vectors  $\bf \Delta=\p_0+\p_1$ and $\k=\frac{1}{2}(\p_0-\p_1)$  for diffractive dijet production in DIS.

In \fig{f-teta}, we show diffractive dijet photo-production correlations $C$ as a function of the angle $\Delta\theta=\theta_{-}-\theta_{+}$ between  vectors $\bf\Delta$ and  $\k$, at a fixed $\Delta\varphi=\pi$ (back-to-back jets) for various values of $A=0,0.5, 1$.  We integrate over the rapidities of two jets $\eta_0, \eta_1\in [0,6]$, $\eta_0, \eta_1\in [6,7.4]$ and also  over the transverse momenta of the two jets within $0.2\leq \p_{0}, \p_{1} [\text{GeV}] \leq 2$. It is seen that a nonzero $A$  corresponding to the existence of $\r-\b$ correlations in the color dipole amplitude, induces sizeable azimuthal correlations
 between  $\bf \Delta$ and  $\k$.

\section*{Acknowledgments}
We thank Cyrille Marquet and Stefan Schmitt for comments on the first version of the manuscript. AHR would like to express his gratitude to Universidade de Santiago de Compostela, for very warm hospitality during his stay when some parts of this work were carried out. This research  was supported by the People Programme (Marie Curie Actions) of the European Union's Seventh Framework Programme FP7/2007-2013/ under REA grant agreement \#318921 (TA, NA and GB);  Fondecyt grants 1150135 and Conicyt grant C14E01, of Chile (AHR); the Kreitman Foundation and the  Israel Science Foundation grant \#87277111 (GB);  the European Research Council grant HotLHC ERC-2011-StG-279579, Ministerio de Ciencia e Innovaci\'on of Spain under project FPA2014-58293-C2-1-P, Xunta de Galicia (Conseller\'{\i}a de Educaci\'on and Conseller\'\i a de Innovaci\'on e Industria - Programa Incite),  the Spanish Consolider-Ingenio 2010 Programme CPAN and  FEDER (TA and NA).


 
\end{document}